\documentclass[10pt,letterpaper]{article}
\usepackage[top=0.85in, left=1in, right=1in, bottom=1in, footskip=0.75in, marginparwidth=2in]{geometry}

\usepackage[utf8]{inputenc}
\usepackage{cite}

\usepackage{amsmath} 
\usepackage{nameref,hyperref}

\usepackage{graphicx}
\usepackage{booktabs}
\usepackage{array}

\usepackage[right]{lineno}

\usepackage{microtype}
\DisableLigatures[f]{encoding = *, family = * }
\usepackage{changepage}
\usepackage[aboveskip=1pt,labelfont=bf,labelsep=period,singlelinecheck=off]{caption}

\makeatletter
\renewcommand{\@biblabel}[1]{\quad#1.}
\makeatother

\usepackage{lastpage,fancyhdr,graphicx,multirow,amssymb}
\usepackage{epstopdf}
\fancyheadoffset[L]{0in} 
\fancyfootoffset[L]{0in}

\usepackage{color}
\definecolor{Gray}{gray}{.25}
\usepackage{graphicx}
\usepackage{sidecap}

\usepackage{wrapfig}
\usepackage[pscoord]{eso-pic}
\usepackage[fulladjust]{marginnote}
\reversemarginpar

\begin{document}
\vspace*{0.35in}

\begin{flushleft}
{\Large
\textbf\newline{PLM-eXplain: Divide and Conquer the Protein Embedding Space}
}
\newline


Jan van Eck \textsuperscript{1,*},
Dea Gogishvili \textsuperscript{1},
Wilson Silva \textsuperscript{1},
and Sanne Abeln \textsuperscript{1}
\\
\bigskip
\bf{1} AI Technology for Life, Department of Computing and Information Sciences, Department of Biology, Utrecht University, Utrecht, Netherlands
\\

\bigskip
*Corresponding author. E-mail: j.vaneck@uu.nl

\end{flushleft}

\section*{Abstract}
Protein language models (PLMs) have revolutionised computational biology through their ability to generate powerful sequence representations for diverse prediction tasks. However, their black-box nature limits biological interpretation and translation to actionable insights. We present an explainable adapter layer - PLM-eXplain (PLM-X), that bridges this gap by factoring PLM embeddings into two components: an interpretable subspace based on established biochemical features, and a residual subspace that preserves the model's predictive power. Using embeddings from ESM2, our adapter incorporates well-established properties, including secondary structure and hydropathy while maintaining high performance. We demonstrate the effectiveness of our approach across three protein-level classification tasks: prediction of extracellular vesicle association, identification of transmembrane helices, and prediction of aggregation propensity. PLM-X enables biological interpretation of model decisions without sacrificing accuracy, offering a generalisable solution for enhancing PLM interpretability across various downstream applications. This work addresses a critical need in computational biology by providing a bridge between powerful deep learning models and actionable biological insights.\\
\\
\textbf{keywords:} Protein language model, Explainable AI, Embeddings, Protein Property Prediction, Protein Aggregation.\\


\section*{Introduction}

The field of computational biology has expanded rapidly, supported by the development of large-scale protein language models (PLMs) trained on extensive sequence databases \cite{Elnaggar2021ProtTrans, lin2023evolutionary}. These models quickly outperformed existing tools across a variety of protein prediction tasks \cite{Elnaggar2021ProtTrans, bepler2021learning, lin2023evolutionary, yu2023enzyme, hie2024efficient, hayes2025simulating}, enabling highly accurate predictions of properties ranging from secondary structure \cite{hoie2022netsurfp, Gogishvili2024PatchProt} and subcellular location \cite{Thumuluri2022DeepLoc2} to protein aggregation \cite{Perez2023AggBERT}. A key innovation behind these PLMs is the transformer architecture \cite{Vaswani2017Attention}, which uses multi-head self-attention mechanisms to process entire protein sequences. This allows the model to learn context-dependent representations for each amino acid. This process captures patterns in the sequence that are stored in a numerical representation. The resulting dense embeddings integrate both local and global sequence information, making them valuable for various downstream tasks.

A critical challenge remains that the representations learned by PLMs are not interpretable, in contrast to traditional shallow learning approaches that rely on hand-engineered features \cite{hou2022ten}. Although traditional methods may be limited in their predictive power, their use of carefully crafted features, such as physicochemical properties and various experimental annotations, provides clear biological meaning to their predictions \cite{waury2024proteome}. PLMs, however, transform protein sequences into high-dimensional amino acid-based representations without specific biological meaning attached, offering limited insight into the underlying principles driving their predictions. This lack of interpretability limits scientific understanding of how PLMs capture biological mechanisms, potentially hindering their integration into experimental workflows where model decisions need clear biological rationales \cite{Vellido2020Interpretability, Frasca2024ExplainableAI}. Efforts to address these issues have focused on internal feature importance, attention-weight analyses \cite{Vig2020Bertology}, and post hoc interpretation methods \cite{Dickinson2022PositionalSHAP}. However, these approaches do not offer a direct mapping between the learned latent representations and biological understanding. As a result, PLMs still lack a bridge between sequence data and biological interpretation.

In this study, we present a new explainable adapter approach, PLM-eXplain (PLM-X) that retains the prediction power of protein language models while including interpretability of the representations. Our method utilises embeddings derived from ESM2 \cite{lin2023evolutionary}, one of the most advanced PLMs currently available. We factored these embeddings into two complementary components: a subspace composed of crafted, interpretable physicochemical features and a compressed residual subspace capturing information not explicitly described by these known attributes. By anchoring a fraction of the embedding space in known descriptors, such as structure classifications (SS3 and SS8) \cite{kabsch1983dictionary} and hydropathy (GRAVY) \cite{kyte1982simple}, we empower researchers to better rationalise the contributions of fundamental chemical and structural factors. The remaining subspace ensures that the model retains its full predictive power, preserving more subtle patterns that contribute to performance but are not captured by the predefined feature set. Our explainable adapter is reusable as a flexible layer that can be integrated into various downstream tasks without requiring retraining the adapter. To illustrate this versatility, we applied our semi-explainable embeddings to three distinct protein-level (global) classification problems: (i) prediction of extracellular vesicle (EV) proteins, which are crucial disease biomarkers occuring in all domains of life \cite{borges2013extracellular, yanez2015biological, vanNiel2018, GamezValero2019}; (ii) identification of transmembrane proteins, a well-characterised task with state-of-the-art solutions \cite{hallgren2022deeptmhmm}; (iii) prediction of protein aggregation propensity (amyloidogenicity), which remains a critical challenge in clinical and biotechnological applications \cite{chiti2017protein, ke2020half, dobson2020amyloid}. In each of these cases, we demonstrate that our explainable adapter not only preserves the high accuracy characteristic of black-box PLM embeddings but also provides a better method to explain the model's decisions.

\section*{Methods}

Our method employs a two-step approach to enhance the interpretability of protein language models while maintaining their predictive power (Figure \ref{model}). First, we train an adapter layer for ESM2 that serves as encoder and transforms traditional PLM embeddings into partitioned representations. This process splits the embedding space into two complementary components: an informed subspace grounded in established biochemical features and a residual subspace that preserves additional predictive information not captured by known attributes. Second, to demonstrate the versatility and effectiveness of these adapted embeddings, we evaluate them across three distinct protein classification tasks: aggregation propensity, EV association, and transmembrane helices classification. For each task, we explore two different architectural approaches: a protein-level (global) analysis method that pools amino acid embeddings by averaging and a local analysis method using convolutional neural networks (CNNs) to capture local patterns of crafted features. 

\subsection*{Partitioning PLM embeddings}

\begin{figure*}[hb!]
\centering
\includegraphics[width=16cm]{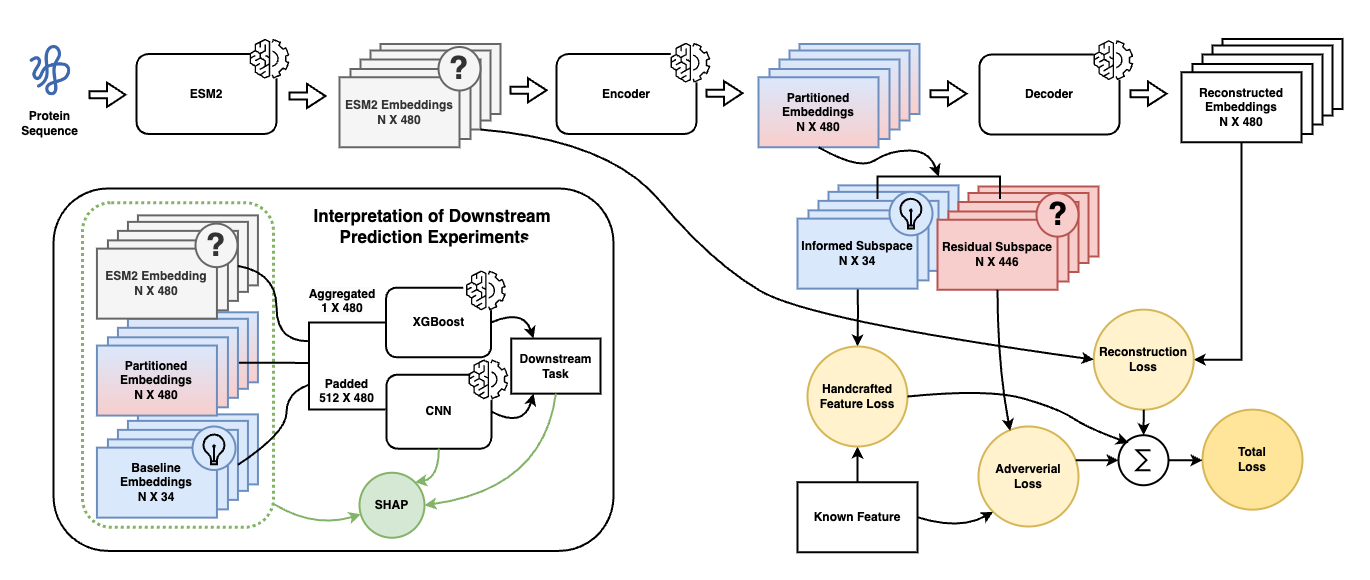}
\caption{\textbf{Our encoder-decoder model architecture} splits protein language model embeddings into two complementary subspaces: one capturing explicit physicochemical features and another containing the residual predictive information. An adversarial component stimulates this separation while preserving the original embeddings' prediction capabilities. We evaluate the partitioned embeddings using XGBoost and CNN models on three downstream protein-level tasks.}\label{model}
\end{figure*}

To transform the ESM2 PLM embeddings (t12\_35M\_UR50D) into partitioned representations, we create two complementary subspaces. The combined embeddings consists of 480 features, matching the size of the original ESM2 embedding. An informed subspace that explicitly captures established biochemical features (N X 34), and a residual subspace (N X 446) that captures the residual predictive information from the original embedding (Figure \ref{model}).

The informed subspace is designed to represent well-understood protein characteristics in a transparent manner. By incorporating (hand)crafted features based on fundamental biochemical properties, this subspace provides direct interpretability for a portion of the model's decision-making process. These features include the following metrics: hydrophobicity scales (GRAVY), aromaticity, secondary structure components (SS3, SS8), accessible surface area (ASA) \cite{miller1987interior} and standard amino acid types (Table \ref{tabsup:crafted_feature_codes} describes crafted features in detail). Each latent feature within this subspace corresponds to a distinct known element.

In contrast, the residual subspace preserves information that cannot be readily explained by biochemical features analysed in this work. This subspace is trained adversarially to minimise redundancy with the knowledge-informed subspace while maintaining the predictive power of the model. This design ensures that subtle patterns and complex relationships in the protein sequence data are not lost in the pursuit of interpretability (Figure \ref{model}).

\subsection*{Model architecture}

To maintain the fidelity of the original embeddings, we employ an auto-encoder architecture with specific optimisation objectives (Figure \ref{model}): \textit{(i)} the encoder transforms ESM2 amino acid embeddings into our partitioned representation, ensuring that handcrafted features are distinctly captured in the informed subspace. \textit{(ii)} Adversarial training is applied to ensure a separation between interpretable and non-interpretable components by minimising the presence of handcrafted features information in the residual subspace. \textit{(iii)} The decoder reconstructs the original PLM embedding from our partitioned representation, ensuring that no essential information from the original embedding is lost during the transformation process. The resulting architecture (Figure \ref{model}) creates a bridge between the powerful partitioned representations learned by the PLMs and the need for biological interpretability, while preserving the full content of information from the original embeddings.

To achieve the partitioned representation, a dual branch architecture is applied to the encoder. The informed subspace branch employs two fully connected layers with 480 and 34 neurons to map the predefined handcrafted features, respectively. The first layer is followed a RELU activation while the last layer is followed by a Tanh function. Each separate prediction task was trained with its own singular scaling parameter to enable accurate predictions for large numerical ranges. The residual subspace branch consists of two fully connected layers, each having 480 and 446 neurons respectively, followed by the same activations as the informed subspace branch. The decoder reconstructs the original PLM embeddings from the concatenated partitioned latent representation. The architecture includes two fully connected layers, each with 480 neurons, followed by RELU activation. Adversarial training is implemented on the residual subspace using task-specific adversarial networks in combination with Gradient Reversal Layers (GRLs) \cite{ganin2016domainadversarialtrainingneuralnetworks}. During the forward pass, the GRL passes embeddings unchanged to the adversarial network. In the backward pass, it scales encoder gradients by a negative factor. 

We evaluated our partitioned (adapted) embeddings with two complementary approaches to capture both global protein properties and local sequence-level features. For global verification, we train a XGBoost classifier with sequence averaged partitioned embeddings on the three different downstream tasks. The XGBoost classifier was configured with 100 estimators, a maximum depth of 5, and a learning rate of 0.1. For the validation of local context, we implemented a convolutional neural network (CNN) architecture. The CNN consisted of a single convolutional layer with 10 filters and a kernel size of 3, 11, or 21 for Aggregation, EV and transmembrane helix prediction respectively. This layer is followed by a ReLU activation and a max pooling operation (MaxPool1D). A single feed forward layer is applied onto the pooled features. The model training process was conducted over a maximum of 40 epochs, with the best-performing model selected based on F1 performance on the validation set. Training was performed with a batch size of 16. For aggregation and transmembrane predictions, a learning rate of 0.001 was applied, while a lower learning rate of 0.0001 was used for EV predictions to optimise performance.

\subsubsection*{Loss functions}

The development of PLM-X is guided by three distinct loss functions, each serving a specific purpose in creating our partitioned embeddings (Figure \ref{model}). Individual loss functions are themselves composites of multiple feature-specific losses, tailored to the nature of each predicted attribute. For multi-class classification tasks, such as secondary structure prediction (SS3 and SS8), we use cross-entropy loss. For binary classification we use binary cross entropy. We used the L1 loss for continuous features, including ASA and GRAVY. 

First, we employ a handcrafted feature loss ($\mathcal{L}_{\text{hcf}}$) that ensures that the informed subspace accurately captures predefined biochemical features. This loss function measures the difference between predicted and actual values of our crafted features, encouraging the model to learn explicit representations of these established protein characteristics. Second, an adversarial feature loss ($\mathcal{L}_{\text{adv}}$) is implemented to maintain the knowledge separation of the two subspaces. Third, a reconstruction loss ($\mathcal{L}_{\text{rec}}$) verifies that the combined information from both subspaces reproduces the original PLM embeddings. This loss function measures the discrepancy between the decoder's output and the initial embeddings, ensuring that no essential information is lost during the transformation process. 

\begin{align}
\mathcal{L}_{\text{total}} = \lambda_{\text{rec}} \mathcal{L}_{\text{rec}}(\mathbf{z}_{\text{orig}}, \mathbf{z}_{\text{recon}}) 
+ \sum_{t=1}^{T} \left( \lambda_{\text{hcf}}^{(t)} \mathcal{L}_{\text{hcf}}^{(t)}(\mathbf{f}_{\text{real}}^{(t)}, \mathbf{f}_{\text{pred}}^{(t)_{\text{hcf}}}) 
+ \lambda_{\text{adv}}^{(t)} \mathcal{L}_{\text{adv}}^{(t)}(\mathbf{f}_{\text{real}}^{(t)}, \mathbf{f}_{\text{pred}}^{(t)_{\text{adv}}}) \right)
\end{align}

The hyperparameter \(\lambda_{\text{rec}}\) represents the weight for the reconstruction loss (\(\mathcal{L}_{\text{rec}}(\mathbf{z}_{\text{orig}}, \mathbf{z}_{\text{recon}})\)), where \(\mathbf{z}_{\text{orig}}\) is the original embedding produced by the encoder, and \(\mathbf{z}_{\text{recon}}\) is the reconstructed embedding. The terms \(\lambda_{\text{hcf}}^{(t)}\) and \(\lambda_{\text{adv}}^{(t)}\) denote the weights for the feature loss (\(\mathcal{L}_{\text{hcf}}^{(t)}\)) and adversarial loss (\(\mathcal{L}_{\text{adv}}^{(t)}\)), respectively, for the \(t\)-th specific task. Here, \(\mathbf{f}_{\text{real}}^{(t)}\) represents the real (ground-truth) feature for task \(t\), and \(\mathbf{f}_{\text{pred}}^{(t)}\) represents the corresponding predicted feature. The term \(T\) corresponds to the total number of tasks. The total loss (\(\mathcal{L}_{\text{total}}\)) is expressed as a weighted sum of the three components, where the hyperparameters \(\lambda_{\text{rec}}\), \(\lambda_{\text{hcf}}^{(t)}\), and \(\lambda_{\text{adv}}^{(t)}\) regulate the contribution of the reconstruction, feature, and adversarial losses, respectively (Figure \ref{model}).

\subsection*{Data collection and curation}

Our model adaptation was performed using 20,298 human protein structures obtained from AlphaFoldDB (accessed December 6, 2024) \cite{jumper2021highly, varadi2024alphafold}. For each amino acid in these protein chains, we calculated multiple structural and physicochemical features using DSSP software \cite{kabsch1983dictionary}, including eight-state secondary structure (SS8), three-state secondary structure (SS3), and ASA (Figure \ref{figsupp:data}). Additional physicochemical properties, including GRAVY scores \cite{kyte1982simple} and aromaticity, were calculated using BioPython \cite{cock2009biopython}. The set of features was completed with the one hot encodings of the 20 standard amino acids (Table \ref{tabsup:crafted_feature_codes}). 

We selected three biologically significant prediction tasks to evaluate our partitioned embeddings: protein aggregation propensity, EV association, and transmembrane helix prediction. These global (protein-level) binary tasks represent diverse challenges in protein sequence analysis, each requiring the detection of distinct physicochemical and structural features, allowing us to assess the biological relevance of the learned representations. For this, we collected three independent datasets. For the EV association protein prediction, we employed a recently curated human proteome dataset \cite{waury2024proteome}. The predictions of protein aggregation propensity were evaluated using the extensively validated amyloid dataset from WALTZ-DB 2.0 \cite{louros2020waltz}. Transmembrane helix proteins were extracted from DeepTMHMM training data. \cite{hallgren2022deeptmhmm}.

\subsection*{Model Interpretation}

The partitioned embeddings were analysed using two complementary methods: SHAP analysis and filter activation-based interpretation. These approaches provided detailed insights into model predictions by quantifying feature importance and exploring the relationship between input sequences and model activations. SHapley Additive exPlanations (SHAP) \cite{lundberg2017unifiedapproachinterpretingmodel} were used to evaluate the contribution of each feature within the partitioned embedding space to model predictions. We employed the TreeExplainer module to compute SHAP values for predictions made by the XGBoost classifier on pooled amino acid embeddings. For local interpretation, we analysed the most activated filter in a 1D CNN trained on amino acid-level embeddings for the transmembrane prediction task, focusing on Leptin (AF-P41159-F1). SHAP values were computed over the input sequences and the activations of this filter, providing insights into the sequence regions most relevant to the model's decisions.

\section*{Results and discussion}

The aim of this work is to introduce an explainable adapter to effectively balance PLM interpretability with predictive power. An initial step of our approach was to train an encoder and transform PLM amino acid embeddings into partitioned embeddings. For this, first, we examined whether the encoder captured handcrafted features distinctly in a respective subspace. In parallel, the residual subspace was adversarially trained to minimise the information about handcrafted features. The decoder reconstructed the original embedding from the partitioned embedding to ensure the conservation of all the information. As a result, the mean absolute reconstruction error (MAE) of 0.068 was achieved, indicating high fidelity in information preservation during the transformation process. To evaluate the effectiveness of our encoding approach, we compared the performance of the knowledge-informed subspace against compressed residual subspace embeddings across multiple protein characteristics (Table \ref{tabsupp:comparison_performance}). The knowledge-informed subspace demonstrated superior performance in all metrics, indicating that the information encoded in the handcrafted features is most strongly encoded in this subspace. 

Note that there is an inherent trade-off between subspace separation and reconstruction accuracy that is tunable in the loss function. Here, we prioritised preserving the performance of the original ESM2, ensuring that our adapted embeddings retain full predictive power while providing interpretable features for downstream analysis. For different scenarios a greater separation may be desirable, for instance, in case of unwanted biases in the data.

\subsection*{Model performance}

Having established the separation of embeddings, we aimed to assess the versatility of PLM-X and the possibility to integrate it into different prediction problems. For this, we evaluated two distinct architectural approaches: a pooled embeddings model in which protein embeddings are averaged, and a CNN model using sliding windows throughout the sequence (Table \ref{tab:pooled_vs_cnn}). These models were tested across three global binary classification tasks: aggregation propensity, EV association, and transmembrane helix prediction. We selected case examples based on both their biological significance and the availability of high-quality curated datasets. 

The CNN architecture demonstrated superior performance in aggregation prediction, achieving a ROC-AUC of 0.90 and F1 score of 0.80 with adapted embeddings, compared to the pooled embeddings model's ROC-AUC of 0.89 and F1 score of 0.76. This suggests that the local sequence context, captured by CNN's sliding window approach, plays an important role in aggregation propensity prediction. Importantly, our method is on par with the state-of-the-art method, AggBERT, which was reported to achieve an AUC of 0.90 on the same dataset \cite{Perez2023AggBERT}. For EV association prediction, the pooled embeddings model slightly outperformed the CNN, achieving a ROC-AUC of 0.79 with adapted embeddings outperforming the previously trained sequence-based classifier \cite{waury2024proteome}. The transmembrane helix prediction showed exceptional performance for both approaches, with both models achieving ROC-AUC values of 0.99 and F1 scores of 0.93 using adapted embeddings. However, for this task, a fair comparison to the state-of-the-art tools is not feasible due to differences in dataset composition (binary versus multiclass) and evaluation protocols. Across all tasks, our adapted embeddings maintained the performance of the original embeddings, while providing interpretable features. It is important to note, that our primary objective was to demonstrate performance parity between the original (ESM2) and our partitioned embeddings, while maintaining comparable accuracy with our crafted only (baseline) model (Table \ref{tab:pooled_vs_cnn}). Although for the downstream prediction tasks the architectural approach was not optimised for pursuing state-of-the-art performance, our models achieved results comparable to recent developments in these tasks. 

\begin{table*}[hb!]
    \centering
    \small
    \caption{Performance comparison between pooled embeddings and CNN across different prediction tasks.}
    \begin{tabular}{l l ccc ccc}
        \hline
        \multirow{2}{*}{\textbf{Prediction task}} & \multirow{2}{*}{\textbf{Embeddings}} & \multicolumn{3}{c}{\textbf{Pooled embeddings}} & \multicolumn{3}{c}{\textbf{CNN}} \\
        \cline{3-5} \cline{6-8}
        & & ROC-AUC & Accuracy & F1 & ROC-AUC & Accuracy & F1 \\
        \hline
        \multirow{3}{*}{Aggregation propensity}
        & Original (ESM2) & 0.90 & 0.84 & 0.76 & 0.90 & 0.85 & 0.80 \\
        & Partitioned (PLM-X) & 0.90 & 0.84 & 0.77 & 0.90 & 0.85 & 0.80 \\
        & Crafted only & 0.89 & 0.83 & 0.74 & 0.89 & 0.83 & 0.76 \\
        \hline
        \multirow{3}{*}{EV association}
        & Original (ESM2) & 0.79 & 0.74 & 0.59 & 0.77 & 0.71 & 0.63 \\
        & Partitioned (PLM-X) & 0.79 & 0.75 & 0.62 & 0.77 & 0.72 & 0.63 \\
        & Crafted only & 0.76 & 0.71 & 0.55 & 0.72 & 0.69 & 0.49 \\
        \hline
        \multirow{3}{*}{Transmembrane helix}
        & Original (ESM2) & 0.99 & 0.98 & 0.92 & 0.99 & 0.99 & 0.95 \\
        & Partitioned (PLM-X) & 0.99 & 0.98 & 0.93 & 0.98 & 0.98 & 0.93 \\
        & Crafted only & 0.97 & 0.97 & 0.87 & 0.97 & 0.96 & 0.90 \\
        \hline
    \end{tabular}
    \label{tab:pooled_vs_cnn}
\end{table*}

\subsection*{Global interpretation}

Our case examples were chosen on the basis of the general understanding of the biological problem and the quality of the available curated datasets. Our main objective was to match the performance between the original ESM2 and the adapted embeddings while regaining the ability from models using crafted sequence features in terms of explainability (Table \ref{tab:pooled_vs_cnn}). 

\begin{figure*}[th!]
\centering
\includegraphics[width=16.5cm]{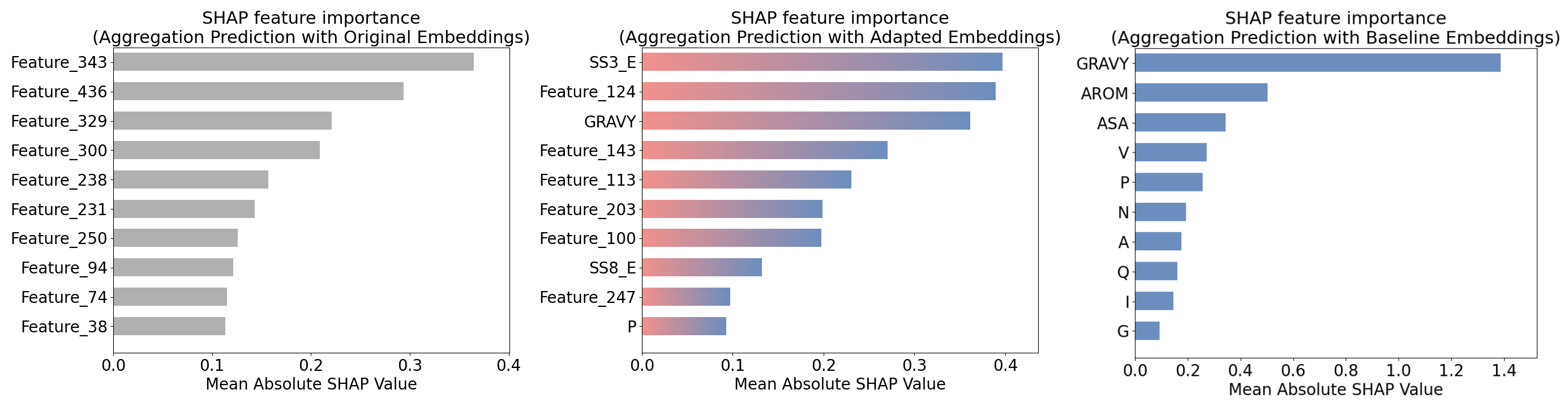}
\caption{\textbf{Global feature importance} for predicting protein aggregation, comparing three different embedding approaches: the original (ESM2) embeddings, partitioned embeddings (PLM-X), and crafted only embeddings. For the original model, top features are unknown. For the adapted model, several known features, such as GRAVY and secondary structure components (extended $\beta$-strands, SS3 E and SS8 E), and proline (P) can be identified. Figure \ref{figsupp:global_shap} shows detailed SHAP plots for all three downstream prediction tasks.}\label{aggregation_global}
\end{figure*}

For global interpretability, the SHAP plots reveal key features driving the predictions for each of our case examples (Figures \ref{aggregation_global} and \ref{figsupp:global_shap}). Although the top features of the original (ESM2) model remain abstract and uninterpretable, our partitioned model identified several biologically relevant properties as important predictors. For the aggregation propensity and transmembrane helix predictions, GRAVY index emerged as a crucial feature along with the secondary structure components - extended $\beta$-strands (SS3 E and SS8 E) agreeing with the known link between amyloidogenicity and hydrophobicity \cite{van2020hydrophobic, thompson2024massive}. The presence of specific amino acids, particularly proline (P), was also identified as significant contributors to the prediction outcome (Figure \ref{aggregation_global}), a finding that aligns with previous findings regarding sequence motifs with negative effects \cite{thompson2024massive}. For predicting protein sorting in EVs, accessible surface area (ASA) along with SS3 C (coil) and SS3 E ($\beta$-strand) were identified in top features (Figure \ref{figsupp:global_shap}). Our findings corroborate previous research \cite{waury2024proteome} which emphasised coil regions and instability index as key factors in EV protein sorting. Similarly, our crafted-only model revealed cysteine (C) as a significant negative predictor, consistent with earlier results \cite{waury2024proteome}. Furthermore, ASA can be interpreted as an indicator of a structural compactness as proteins with large, solvent-exposed regions are generally less stable and more prone to disorder. The observation that ASA is a negative predictor of EV association (Figure \ref{figsupp:global_shap}B) agrees with previous findings that EV proteins are relatively stable and well-structured \cite{waury2024proteome}.

Hence, from the partitioned embeddings (PLM-X), we can learn to what extent the high performing model is based on currently understood biophysical properties. In the case of aggregation prediction, we can conclude that the signal is dominated by beta-strand propensity, and hydrophobicity, as well as the residual embedding (Feature 124 in Figure \ref{aggregation_global}).

\subsection*{Local interpretation}

\begin{figure*}[ht!]
\centering
\includegraphics[width=17cm]{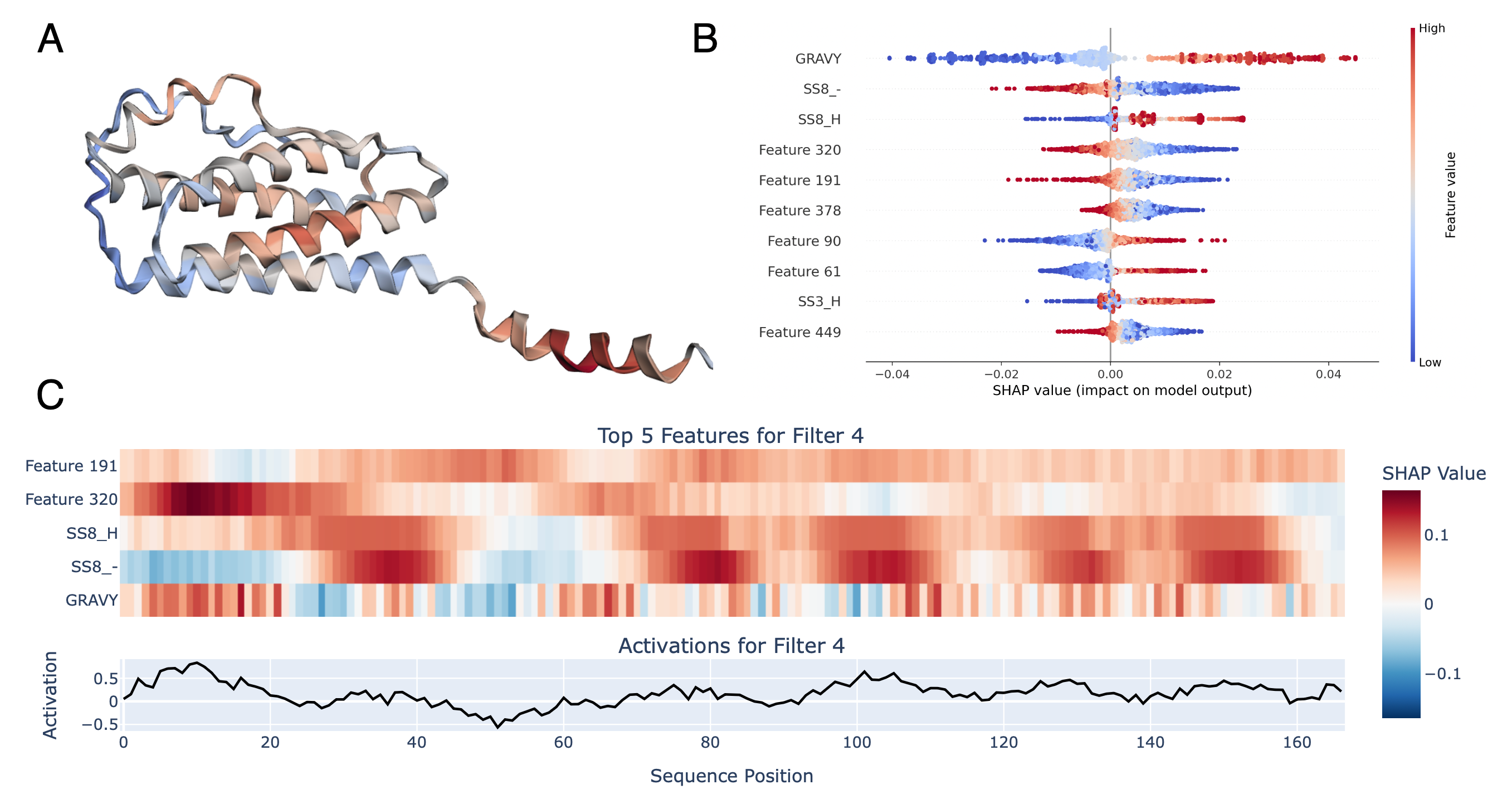}
\caption{\textbf{Local interpretation for the transmembrane helix predictions for the Leptin protein.} \textbf{(A)} The full-length structure of leptin (AF-P41159-F1), highlighted regions are colour-coded based on the highest activation by Kernel 4. \textbf{(B)} The most informative features determined by the sum of absolute SHAP values. Each dot represents a feature at a specific position within a motif. \textbf{(C)} Summed SHAP values for a filter, showcasing the top 5 features at each position. This plot highlights the positional importance of features along the activation values. }\label{local_results}
\end{figure*}

The CNN architecture enables detailed analysis of amino acid-level features and sequence motifs, providing deeper insights into how specific sequence patterns and local structural elements influence the model's predictions. The transmembrane prediction of Leptin was analysed to understand the contributions of specific features to the predictions made by our model (Figure \ref{local_results}A-C). For the most activated kernel (Kernel 4), SHAP values were calculated on the partitioned embeddings, providing insights into feature importances (Figure \ref{local_results}B). This analysis revealed that the GRAVY index, disorder, and alpha-helix features were the most informative predictors across the sequence. An unidentified feature (Feature 320) emerged as highly informative, influencing predictions for the N-terminal region of the protein. This most informative features align with the presence of transmembrane alpha-helix, suggesting that Feature 320 may capture structural or biochemical properties related to such regions. This demonstrates how our approach can validate existing understanding while potentially uncovering new insights.

This study set out to introduce \textit{"the best of both worlds"}: an explainable adapter - PLM-X, that effectively balances PLM interpretability with predictive power. Our approach to overlay established biological knowledge represented through crafted feature explanations onto the complex relationships captured in PLMs' high-dimensional embeddings enables us to differentiate between predictions that align with known biological principles and those that potentially reveal novel mechanistic insights not previously characterised through traditional approaches. The robustness of the architecture in maintaining high performance while providing interpretable features across diverse prediction tasks demonstrates its potential as a general-purpose tool for protein property prediction analysis.

\section*{Conclusion and future outlook}

In this study, we present an innovative approach for interpreting protein language models. We used existing ESM2 embeddings and factored them into two complementary components: a subspace composed of hand-crafted, interpretable features and a compressed residual subspace capturing information not explicitly described by these known attributes. To demonstrate our use-cases, we applied our partitioned embeddings to three distinct classification problems and explained model predictions both on amino acid and protein levels. We showed that our explainable adapter predicts with high accuracy and most importantly provides a possibility to explain the decision of the model. Our explainable adapter provides a versatile foundation that can be applied to a wide range of downstream prediction tasks. PLM-X can be used without requiring any retraining of the ESM2 model or the adapter. The embeddings generated by the PLM-X adapter can be directly applied onto any downstream task, regardless of the type of machine learning model architecture.

This study addresses a fundamental challenge in computational biology, the trade-off between model performance and interpretability. By maintaining high prediction accuracy while providing meaningful biological insights, PLM-X offers a promising direction for developing more trustworthy and actionable AI tools in biological research. Future work could usefully explore the integration of additional physicochemical features and structural information. For scenarios where latent features emerge as significant predictors (e.g., feature 320 in transmembrane helix prediction), systematic correlation analysis with known biological properties could reveal new insights. Following approaches such as sparse auto encoders \cite{simon2024interplm}, could help identify whether these features represent novel biological concepts or combinations of known properties in superposition. This is particularly valuable for expanding our understanding of how PLMs encode biological information and potentially discovering new protein motifs or structural patterns. 

\section*{Data availability statement}
Data and code related to this work can be obtained on request to replicate results from the corresponding author.

\section*{Competing interests}
Outside the submitted work: SA reports a patent pending; SA is in a consortium agreement with Cergentis BV as part of the TargetSV project; SA is in a consortium agreement with Olink and Quanterix as part of the NORMAL project. The rest of the authors do not have competing interests to declare.

\section*{Author contributions statement}
Conceptualisation: JvE; SA; WS;
Data collection: JvE, DG;
Data curation: JvE, DG;
Methodology: JvE;
Formal analysis: JvE;
Funding acquisition: SA;
Supervision: WS, SA;
Visualisation: JvE, DG;
Writing - original draft preparation: JvE, DG, WS, SA;
Writing - review \& editing: JvE, DG, WS, SA.

\bibliographystyle{ieeetr}
\bibliography{references}

\appendix

\renewcommand{\thefigure}{S\arabic{figure}}
\renewcommand{\theequation}{S\arabic{equation}}
\renewcommand{\thetable}{S\arabic{table}}
\setcounter{figure}{0}
\setcounter{equation}{0}
\setcounter{table}{0}

\newpage
\section*{Supporting Information}

\subsection*{Supplementary tables}

\begin{table}[h!]
\centering
\begin{tabular}{|c|p{10cm}|}
\hline
\textbf{Crafted Features} & \textbf{Description} \\
\hline
ASA       & Accessible Surface Area. \\
\hline
SS8\_H    & Alpha-helix in 8-class secondary structure prediction. \\
\hline
SS8\_E    & Beta-strand in 8-class secondary structure prediction. \\
\hline
SS8\_G    & 3-10 helix in 8-class secondary structure prediction. \\
\hline
SS8\_I    & Pi-helix in 8-class secondary structure prediction. \\
\hline
SS8\_B    & Beta-bridge in 8-class secondary structure prediction. \\
\hline
SS8\_T    & Turn in 8-class secondary structure prediction. \\
\hline
SS8\_S    & Bend in 8-class secondary structure prediction. \\
\hline
SS8\_-    & Coil in 8-class secondary structure prediction. \\
\hline
SS3\_H    & Alpha-helix in 3-class secondary structure prediction. \\
\hline
SS3\_E    & Beta-strand in 3-class secondary structure prediction. \\
\hline
SS3\_C    & Coil in 3-class secondary structure prediction. \\
\hline
A         & Amino acid: Alanine. \\
\hline
C         & Amino acid: Cysteine. \\
\hline
D         & Amino acid: Aspartic Acid. \\
\hline
E         & Amino acid: Glutamic Acid. \\
\hline
F         & Amino acid: Phenylalanine. \\
\hline
G         & Amino acid: Glycine. \\
\hline
H         & Amino acid: Histidine. \\
\hline
I         & Amino acid: Isoleucine. \\
\hline
K         & Amino acid: Lysine. \\
\hline
L         & Amino acid: Leucine. \\
\hline
M         & Amino acid: Methionine. \\
\hline
N         & Amino acid: Asparagine. \\
\hline
P         & Amino acid: Proline. \\
\hline
Q         & Amino acid: Glutamine. \\
\hline
R         & Amino acid: Arginine. \\
\hline
S         & Amino acid: Serine. \\
\hline
T         & Amino acid: Threonine. \\
\hline
V         & Amino acid: Valine. \\
\hline
W         & Amino acid: Tryptophan. \\
\hline
Y         & Amino acid: Tyrosine. \\
\hline
GRAVY     & Grand Average of Hydropathy (measure of hydrophobicity). \\
\hline
AROM      & Aromaticity (frequency of aromatic amino acids). \\
\hline
\end{tabular}
\caption{Crafted feature codes and their descriptions.}
\label{tabsup:crafted_feature_codes}
\end{table}

\begin{table}[ht]
    \centering
    \caption{Comparison of performance between the crafted subspace and compressed residual subspace embeddings. ASA, accessible surface area; SS8, eight-state secondary structure; SS3, three-state secondary structure; AA, amino acid type; GRAVY, hydropathy index; Arom, aromaticity; Acc, accuracy.}
    \begin{tabular}{lcccccc}
        \hline
        \textbf{Subspace} & \textbf{ASA ($R^2$)} & \textbf{SS8} (Acc) & \textbf{SS3} (Acc) & \textbf{AA} (Acc) & \textbf{GRAVY} ($R^2$) & \textbf{Arom} (Acc) \\
        \hline
        Knowledge-informed & 0.70 & 0.65 & 0.85 & 1.00 & 0.99 & 1.00 \\
        Compressed residual & 0.67 & 0.64 & 0.83 & 0.88 & 0.89 & 0.98 \\
        \hline
    \end{tabular}
    \label{tabsupp:comparison_performance}
\end{table}

\subsection*{Supplementary figures}

\begin{figure*}[th!]
\centering
\includegraphics[width=15cm]{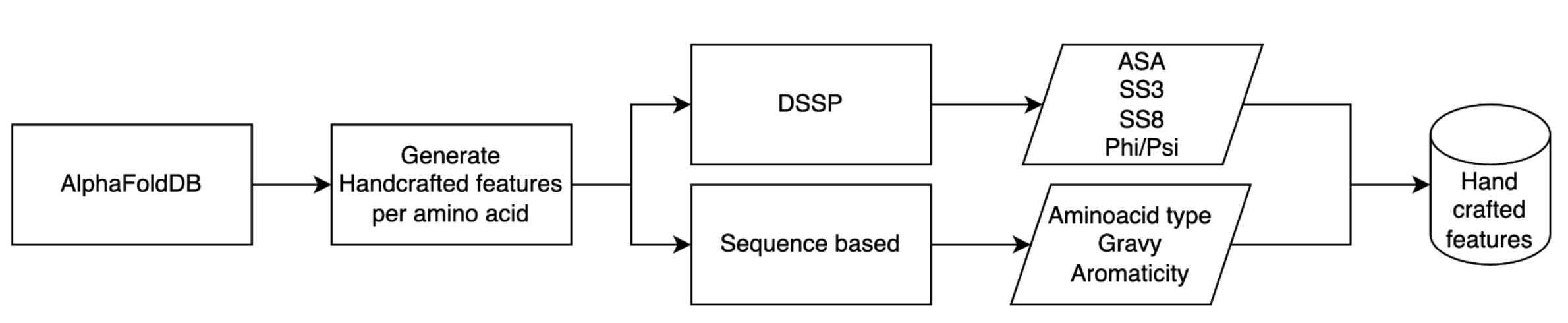}
\caption{\textbf{Data curation pipeline for the model adaptation.} Human proteome from AlphaFoldDB \cite{jumper2021highly, varadi2024alphafold} was annotated with secondary structure components and other sequence-based features. Resulting 34 features were used to create knowledge informed subspace.}\label{figsupp:data}
\end{figure*}

\begin{figure*}[th!]
\centering
\includegraphics[width=16.5cm]{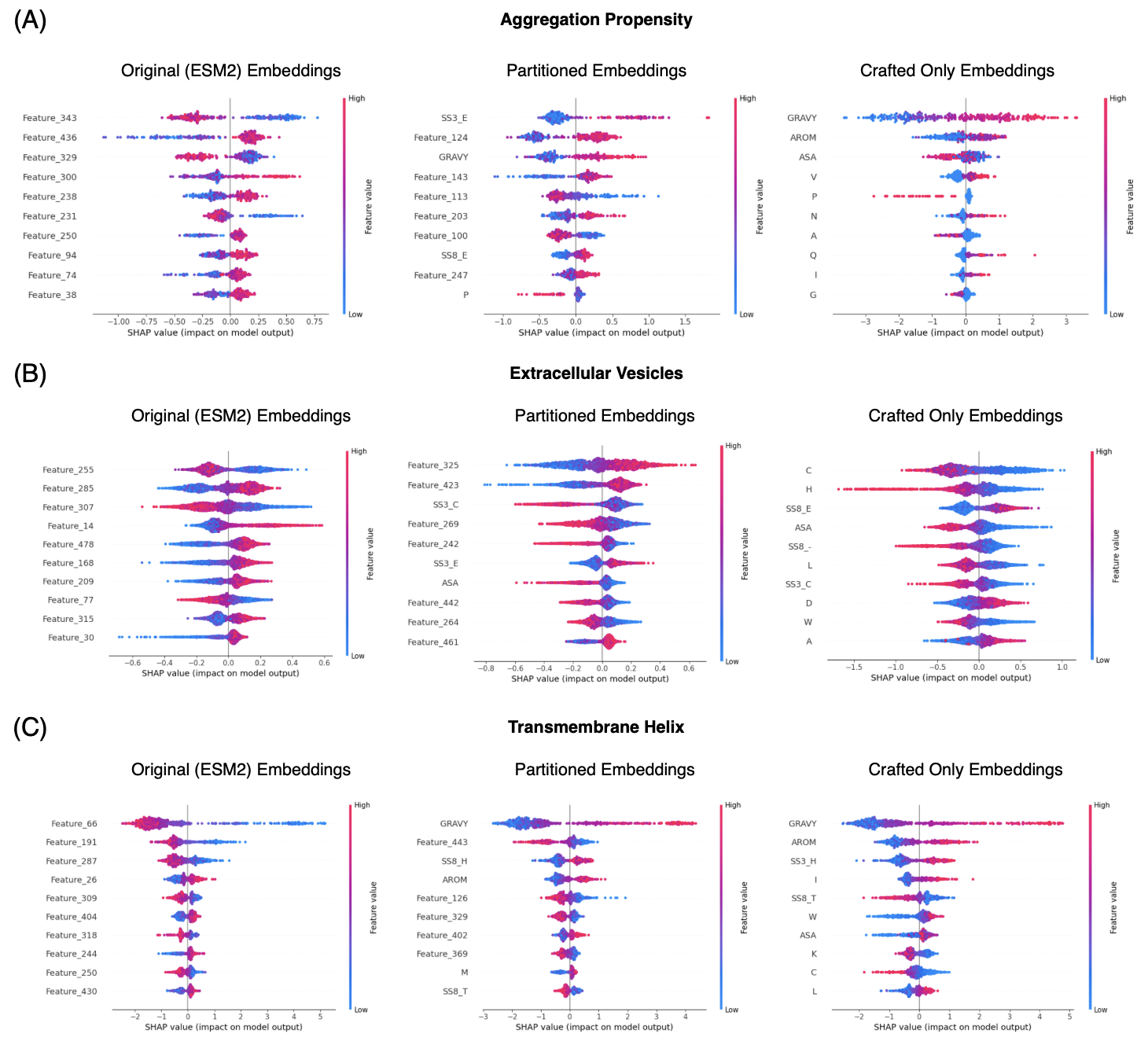}
\caption{\textbf{SHAP summary plots for global interpretability} for three different downstream prediction tasks: \textbf{(A)} aggregation propensity prediction, \textbf{(B)} association with extracellular vesicles (EV) and \textbf{(C)} transmembrane helix predictions. For each prediction task feature importances are shown for the original (ESM2), partitioned, and crafted only (baseline) embeddings. The plot shows a summary of how the top features in a dataset impact the model’s output. Each instance of the explanation is represented by a single dot on each feature row. Colour is used to display the original value of a feature. For the original model, top features are unknown. For the partitioned model, several known features, such as secondary structure features (SS8, SS3), GRAVY, and accessible surface area (ASA) are displayed. For the baseline model, which only uses knowledge-informed subspace of embeddings, all the features are explainable.}\label{figsupp:global_shap}
\end{figure*}

\end{document}